\documentclass{iaus}
\usepackage{graphicx}

\title[IAUS 265.~~Fe-peak element abundances in disk and halo stars] 
{Fe-peak element abundances \\in disk and halo stars}

\author[Maria Bergemann \& Thomas Gehren]
{Maria Bergemann$^1$ \and Thomas Gehren$^2$}

\affiliation{$^1$Max-Planck Institute for Astrophysics,
Karl-Schwarzschildt str. 1\\ 87541, Garching, Germany \\ email:
{\tt mbergema@mpa-garching.mpg.de} \\[\affilskip]
$^2$University Observatory Munich, Scheiner str. 1\\ 81679, Munich, Germany
\\email: {\tt thomas.gehren@gmx.de}}

\pubyear{2009}
\volume{265}
\jname{Chemical Abundances in the Universe: Connecting First Stars to Planets}
\editors{K. Cunha, M. Spite \& B. Barbuy, eds.}
\begin{document}

\maketitle

\begin{abstract}
At present none of galactic chemical evolution (GCE) models provides a
self-consistent description of observed trends for all iron-peak elements with
metallicity simultaneously. The question is whether the discrepancy is due to
deficiencies of GCE models, such as stellar yields, or due to erroneous
spectroscopically-determined abundances of these elements in metal-poor stars.
The present work aims at a critical reevaluation of the abundance trends for
several odd and even-$Z$ Fe-peak elements, which are important for understanding
explosive nucleosynthesis in supernovae.
\end{abstract}

As a rule, abundances of Fe-peak elements in the atmospheres of the Sun and
metal-poor stars are calculated using a local thermodynamic equilibrium (LTE)
assumption for the element line formation. However, recent studies indicate that
LTE breaks down for majority of metals in stellar atmospheres, in particular,
for neutral atoms with low ionization potential.
Our earlier investigations of Mn and Co in cool stars (\cite[Bergemann \& Gehren
2007, 2008, Bergemann et al. 2009]{Bergemann08, Bergemann09}) confirm large
departures from LTE in the excitation-ionization balance of these elements,
which are mainly stipulated by overionization of neutral atoms. As a result, the
LTE-based abundances of Mn and Co in metal-poor stars are severely
underestimated. Under NLTE, the trend of [Mn/Fe] with [Fe/H] is only slightly
subsolar, whereas [Co/Fe] ratios steadily increase with decreasing Fe abundances
in the disk and halo stars.

\begin{figure}[hb]
\begin{center}
\hbox{\includegraphics[width=2.7in]{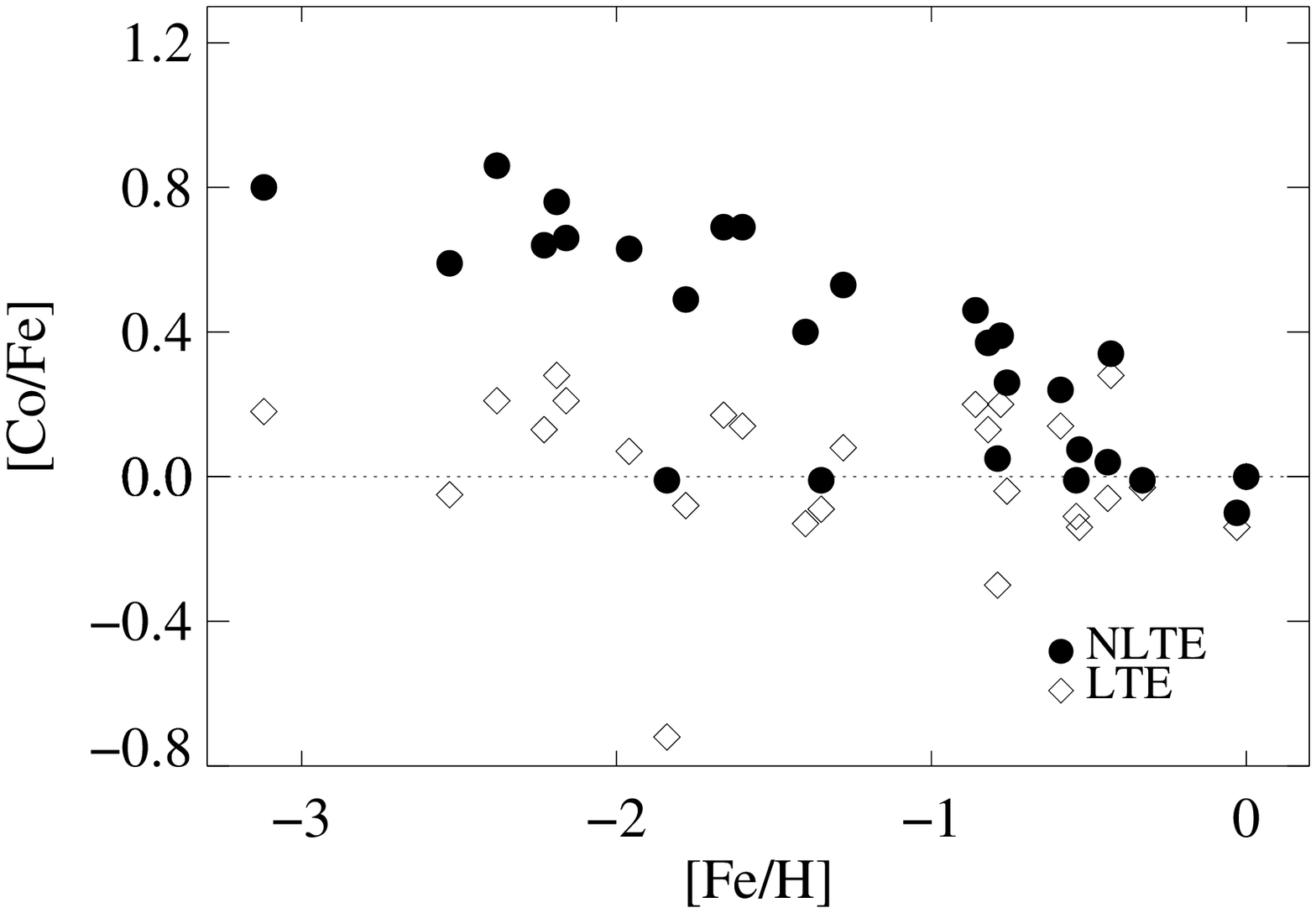}\hfill
\includegraphics[width=2.7in]{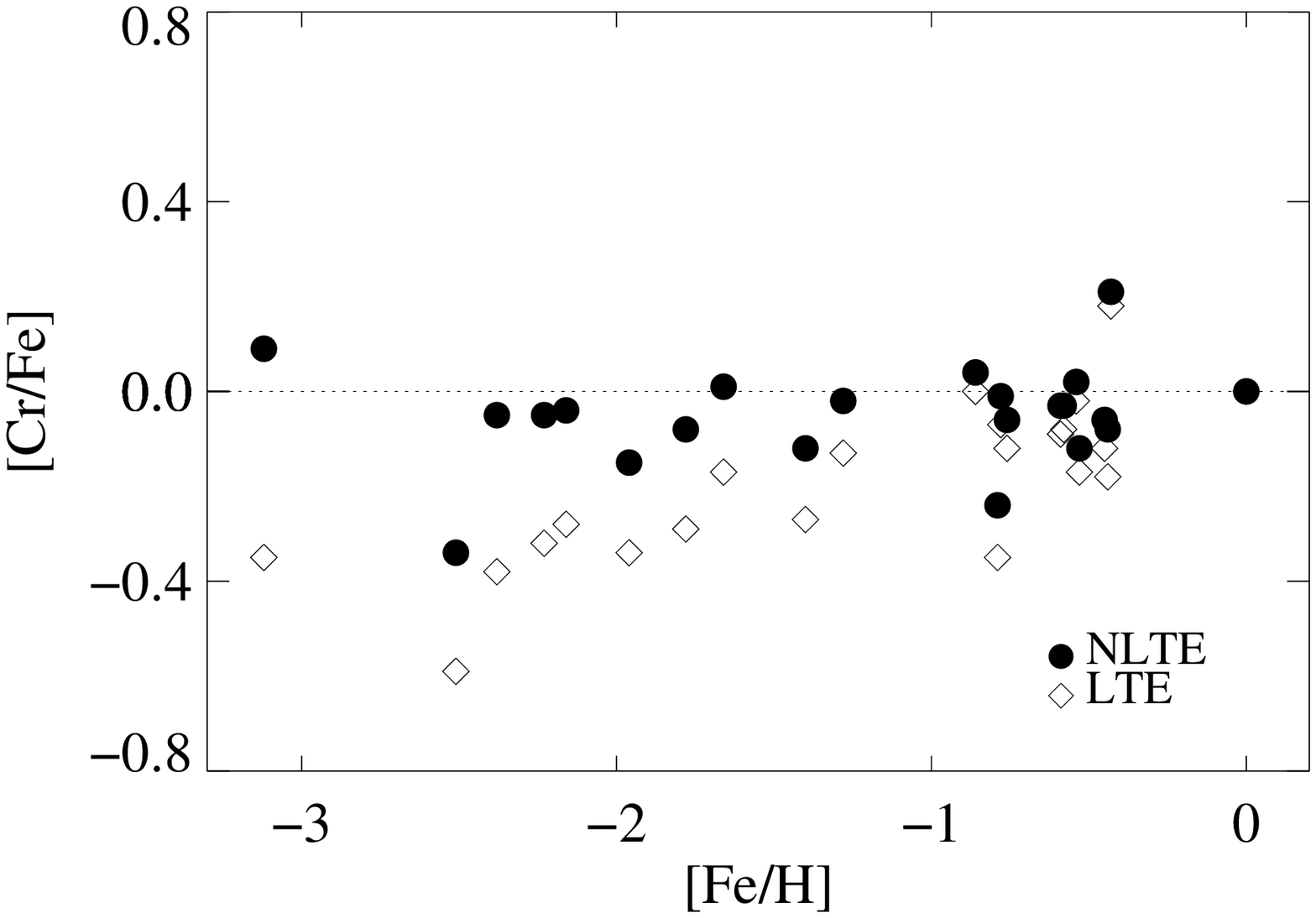}}
 \caption{[Co/Fe] and [Cr/Fe] ratios in metal-poor stars as a function of
metallicity. The abundances of Co and Cr are determined from the lines of
neutral atoms.}
\label{fig1}
\end{center}
\end{figure}

In this work, we have expanded our sample of disk and halo stars from
\cite[Bergemann et al. (2009)]{Bergemann09} with $9$ members of the thin and
thick disks. Contrary to the prevalent belief that NLTE effects on differential
abundances are minor for less metal-poor stars, we find that NLTE abundance
corrections for Co are significant even at [Fe/H] $\sim -0.8$,
$\log\varepsilon^{\rm NLTE} - \log\varepsilon^{\rm LTE} \sim +0.25$ dex. Our new
data reveal a well-defined increase of [Co/Fe] ratios at [Fe/H] $\sim -0.5$
(Fig.\,\ref{fig1}). An ostensive dichotomy of [Co/Fe] values seen for the stars
with $-1 <$ [Fe/H] $<-0.5$ is not significant. Apparently, 
there is a large continuous spread of Co abundances in stars with mildly
sub-solar metallicities, which can be explained by GCE models with radial
migration (\cite[Sch\"{o}nrich \& Binney 2009]{Ralph09}).

We find similar NLTE effects on abundances of Cr in metal-poor stars. The
details of statistical equilibrium calculations will be published elsewhere.
We only note that quantum-mechanical photoionization cross-sections for Cr I
are now available \cite[(Nahar 2009)]{Nahar09}. This allowed us to
put constraints on poorely-known cross-sections for inelastic collisions with H I, which are computed according to \cite[Drawin (1969)]{Drawin69} and scaled down by few orders of magnitude. Under
NLTE, ionization equilibrium in Cr I/Cr II is shifted to lower number densities
of Cr I, thus requiring larger Cr abundances to fit the observed spectral lines
of Cr I. The main stellar parameter that determines the sign and magnitude of
NLTE abundance corrections is metallicity: the effect of overionization on the
line opacity monotonously increases with decreasing [Fe/H]. In addition to the
metallicity effect, high temperatures control overionization at higher [Fe/H],
whereas at low [Fe/H] the effect of low gravity is more important. Hence, NLTE
line formation and abundance corrections for giants and dwarfs at a given [Fe/H]
are different for transitions involving different levels of Cr I.

NLTE abundances of Cr were calculated for a sample of dwarfs and subgiants
(Fig.\,\ref{fig1}) with atmospheric parameters taken from \cite[Bergemann et al.
(2009)]{Bergemann09}. In short, the effective temperatures and surface gravities
are determined from Balmer line profiles and \textit{hipparcos} parallaxes,
respectively. The iron abundances and microturbulent velocities were obtained
from Fe II line profile fitting under LTE requiring that the derived Fe
abundances are independent of the line strength. Ionization equilibrium of Cr in
metal-poor stars is satisfied when we apply a very small scaling factor to 
cross-sections for inelastic collisions with H I, S$_{\rm H} \leq 0.05$. With
this choice of S$_{\rm H}$, our study yields [Cr/Fe] $\sim 0$ throughout the
range of metallicities analyzed here, $-3 \leq$ [Fe/H] $\leq 0$. The
metallicity-independent [Cr/Fe] ratios in metal-poor stars are well reproduced
by most of the GCE models (e.g. \cite[Samland 1998, Kobayashi et al. 2006]
{Samland98, Kobayashi06}). On the other side, these same models tend to
underestimate the NLTE [Mn/Fe] and [Co/Fe] ratios in metal-poor stars.
It is tempting to relate the discrepancy to stellar yields used in the GCE
models. According to \cite[Kobayashi et al. (2006)]{Kobayashi06}, it is possible
to find a combination of SN II explosion parameters, like mass cut and amount
of mixing, to reproduce [Cr/Fe], [Mn/Fe], and [Co/Fe] simultaneously. This
problem will be investigated further.

\end{document}